\documentclass[5p,sort,compress]{elsarticle}
\usepackage{amsfonts}
\usepackage{amsmath}
\usepackage[labelfont=bf,justification=raggedright,singlelinecheck=false]{caption}
\usepackage{array}
\usepackage{tabu}
\usepackage{multirow}
\usepackage{graphicx}
\usepackage{amsmath}
\usepackage{subcaption}
\usepackage[noend]{algpseudocode}
\usepackage[linesnumbered,ruled,vlined]{algorithm2e}
\usepackage{pdflscape}
\usepackage{rotating}
\usepackage{wrapfig}
\usepackage[inline, shortlabels]{enumitem}
\usepackage{mathtools}
\usepackage{amsmath}

\newcolumntype{v}{>{\centering\arraybackslash}m{.18\linewidth} }
\begin{document}
\journal{Visual Communication Image Representation}
\title{A blind robust watermarking method based on Arnold Cat map and amplified pseudo-noise strings with weak correlation}
\author[fum]{Seyyed Hossein Soleymani}
\ead{seyyedhosein.soleymani@mail.um.ac.ir}
\author[fum]{Amir Hossein Taherinia\corref{cor1}}
\ead{taherinia@um.ac.ir}
\author[fum]{Amir Hossein Mohajerzadeh}
\ead{mohajerzadeh@um.ac.ir}
\cortext[cor1]{Corresponding author}
\address[fum]{Computer Engineering Department, Ferdowsi University of Mashhad, Mashhad, Iran }
\begin{abstract}
In this paper, a robust and blind watermarking method is proposed, which is highly resistant to the common image watermarking attacks, such as noises, compression, and image quality enhancement processing. In this method, Arnold Cat map is used as a pre-processing on the host image, which increases the security and imperceptibility of embedding watermark bits with a strong gain factor. Moreover, two pseudo-noise strings with weak correlation are used as the symbol of each 0 or 1 bit of the watermark, which increases the accuracy in detecting the state of watermark bits at extraction phase in comparison to using two random pseudo-noise strings. In this method, to increase the robustness and further imperceptibility of the embedding, the Arnold Cat mapped image is subjected to non-overlapping blocking, and then the high frequency coefficients of the approximation sub-band of the FDCuT transform are used as the embedding location for each block. Comparison of the proposed method with recent robust methods under the same experimental conditions indicates the superiority of the proposed method.
\end{abstract}
\maketitle
\section{Introduction}
\noindent Hiding information in image has been widely used in recent years. The numerous applications of hiding information in today's life have made this science be divided into subcategories, based on the application. The two main subcategories of information hiding are watermarking and steganography. A common important point in these subcategories is the hiddenness of the message, watermark, data, or information. In steganography methods, the goal is to transfer the information through an image securely and invisibly, so that embedding does not create tangible changes in the image and is not recognizable through steganalysis methods that attempt to detect whether carrying the message or not by an image~\cite{ref001,ref002,ref003}.

Today, with the growth of virtual communications and the dramatic advance of the Internet, access to text, audio, image and video is provided to everyone, and protecting these documents against the forgery, manipulation, and violation of property rights is a requirement of such a space. Without losing the whole subject, using encryption methods and digital signature, changes made in the digital image can be found, however, this only works until the unintentional attacks, such as compression, image quality processing, poor noise, etc., cause no change in the image; since the digital signature accompanying the image will not be the same as what is extracted by the receiver. In this case, it can only be determined whether the image has changed or not, and the intentional or unintentional attack cannot be determined.

The watermarking can be divided based on the amount of information required to extract the watermark in the destination into three blind, semi-blind, and non-blind categories. In the blind watermarking, the main signal is not required during the extraction process, and only the keys are needed. In other words, in these methods, the unembedded original image is not needed to extract the watermark, and only the watermarked image and a few simple keys are needed~\cite{ref4,ref9}. In semi-blind watermarking, in some cases, we need additional information to extract the watermark. For example, in some SVD-based methods, to extract the embedded watermark in the destination additional information (such as eigenvectors of the original image) is required~\cite{ref6,ref10,ref11,ref12}. In the non-blind watermarking, the original image is required to extract the watermark. These methods are generally more robust than blind watermarking methods, instead, are not very common and functional due to the need to send additional information~\cite{ref13,ref14,ref15}.

In terms of embedding domain, the watermarking is divided into two categories of spatial domain and transform domain. The spatial domain-based methods spread the watermark data in pixel values of the original image and creates a very small change in image brightness. This methods have less computational complexity and do not require a specific transform. The spatial domain-based methods, although performs well in terms of imperceptibility, operate poorly in terms of robustness to signal processing attacks such as image compression, low pass filtering and noises. The simplest method of this domain is embedding the watermark in the least significant bits of image pixels~\cite{ref16,ref17}. To provide simultaneous imperceptibility and robustness, the watermark is embedded in the transform domain of cover image. In this method, coefficients of transformed image are changed to embed the watermark. The domain of the transform is also called the frequency domain; since the signal is changed from its original form and decomposed into frequency components. The most important and widely used methods in the frequency domain are the discrete cosine transform (DCT), the discrete wavelet transform (DWT), the integer wavelet transform (IWT) and other transforms of the X-Let family~\cite{ref18,ref19,ref20}.

In order to protect the property rights of an image, it is necessary to insert the watermark in the image in a robust manner that is resistant to unintentional attacks such as image compression, adding noise, resizing, and even cutting some part of the image. In this paper, a blind watermarking based on Arnold Cat map method, a fast discrete curvelet transform (FDCuT), and a DCT transform is proposed which uses two random pseudo-noise with a low correlation as the symbol of the bits 0 and 1 of the watermark. This method has the high robustness to intentional and unintentional manipulation, such as intense noises, image compression, and image quality enhancements processing. The rest of the structure of this article is such that in Section 2, related works was investigated. In Section 3, the proposed method is described in detail. In sections 4 and 5, the results of the experiments and conclusion are presented, respectively.

\section{Related Work}
\noindent In this section, a number of robust watermarking methods compared in the evaluation section of the proposed method are briefly described and their main ideas are expressed.

In~\cite{ref7}, an LWT-based blind robust watermarking method was proposed that uses the quantization method of the LH3 sub-band coefficients for embedding. In this method, the coefficients in the LH3 sub-band are subjected to disturbance and then grouped into blocks. The coefficients in each block are arranged, and then two values as the difference between the two maximum and the difference of the two minimum coefficients are obtained. If the difference between the two minimum coefficients is lower than the τ threshold, then the block is considered as an embeddable block. Moreover, to embed in each embeddable block, quantization of the maximum value of that block is used.

In~\cite{ref8}, a blind robust watermarking based on quantization through dither modulation has been proposed which, in addition to optimal robustness, maintains the visual quality of the image. In this method, three levels of discrete wavelet transform are performed on the host image, and then the LH3 and HL3 coefficients are selected and grouped. In this method, the difference between the two minimum and maximum values in each group is calculated and used in the quantization process.

In~\cite{ref4}, a robust blind watermarking method based on BCH error correction coding and the Spread spectrum method is proposed. In this method, the second level approximation sub-band of the discrete wavelet transform was used as the embedding location of the watermark. In this method, the LL2 sub-band are divided into non-overlapping blocks and the high frequency coefficients of each block are selected for embedding the watermark. In this method, Spread Spectrum technique is used as embedding method. Moreover, the watermark bits are encoded by BCH error correction codding before embedding. The robustness of this method against attacks, especially compression, is significant.

In~\cite{ref6}, a semi-blind watermarking method is proposed based on the DWT transform and singular values decomposition (SVD). In this way, in order to increase security, the watermark is first encrypted with a public key and RSA algorithm. Then a DWT transform level is calculated from the host image and the approximate sub-band is selected for embedding. The encoded watermark information is embedded in the eigen-values of LL1 approximation sub-band of cover image. The robustness and the image quality of this watermarking method are high, however, there is the point that it is a semi-blind method, and in order to extract the watermark, the right and left matrices (U, V) obtained from SVD decomposition of the host image are required, which is a weakness for this method.

\section{Proposed method}
\label{sec:rroposedmethod}
\begin{figure*}
  \center
  \includegraphics[width=1\textwidth]{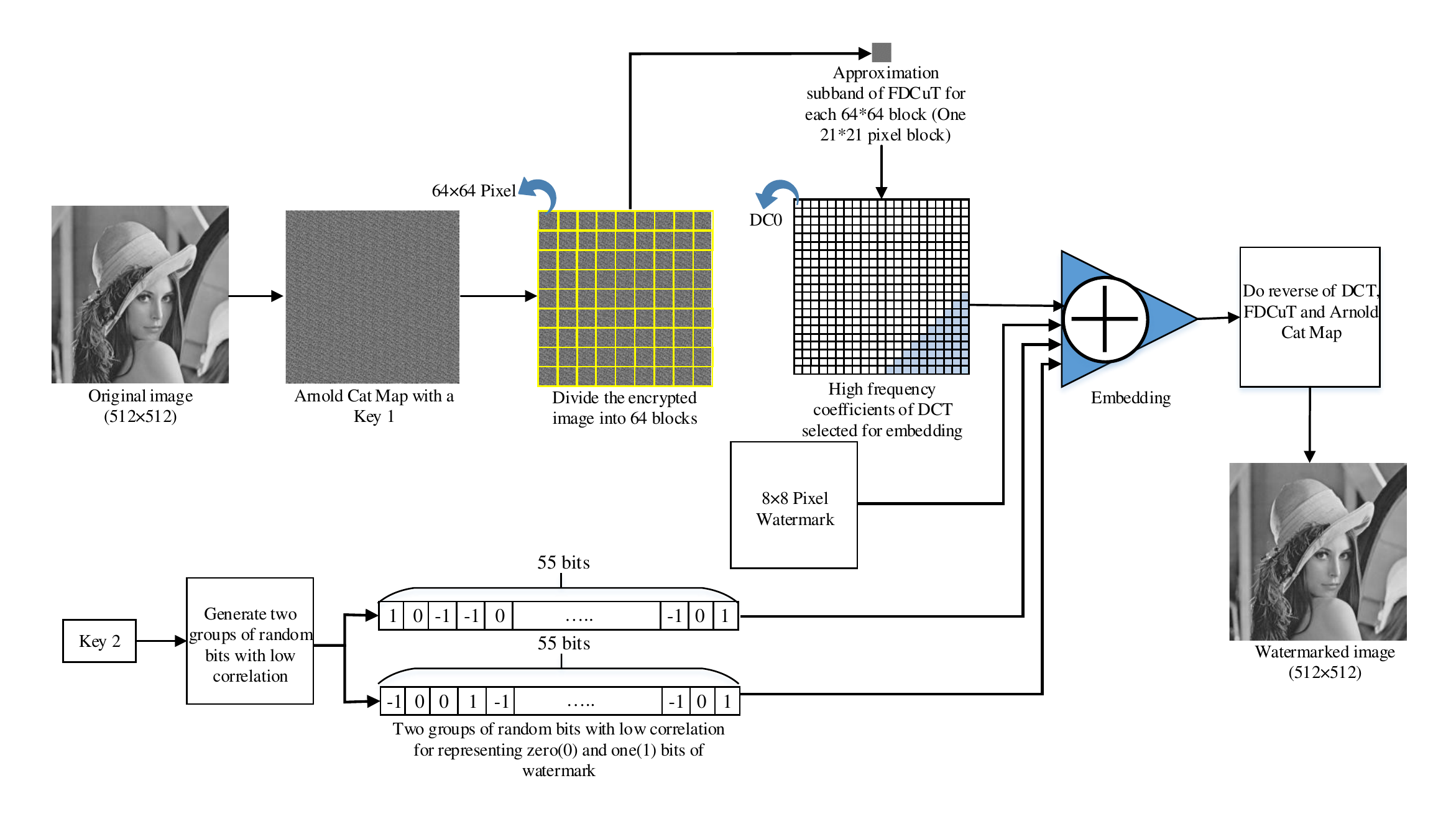}
  \caption{Embedding diagram}\label{fig.embeddingflowchart}
\end{figure*}

\noindent The proposed algorithm involves the advantages of several methods such as Arnold Cat map, FDCuT transform, DCT transform, and two weak correlated noises. Each method and using trend in the proposed method are presented in sub-sections~\ref{subsec.ACM},~\ref{subsec.FDCuT},~\ref{subsec:DCT} and~\ref{subsec:weakcorr}, respectively. The embedding and extraction way will be expressed in sub-sections~\ref{subsec:embedding} and~\ref{subsec.extraction}, respectively. The flowchart of the embedding operation of the proposed method is shown in Fig.~\ref{fig.embeddingflowchart}.

\subsection{Arnold Cat Map}\label{subsec.ACM}

\noindent Arnold Cat map is a two-dimensional mapping, and when applied to a digital image, it changes the original location of the pixels randomly. Arnold Cat map is one of the widely used image processing transforms in the field of encryption and watermarking. This map is a simple, periodic, and reversible transform. The periodicy and reversibility of a transform mean that if we apply a transform successively to a given matrix, then the initial data will be obtain after a complete period. This map is defined as the equation~\ref{eq:arnoldcatmap}

\begin{equation}\label{eq:arnoldcatmap}
\begin{bmatrix} 
x^{'} \\
y^{'} 
\end{bmatrix} = \begin{bmatrix} 
1 & a \\
b & ab+1 
\end{bmatrix}\begin{bmatrix} 
x \\
y 
\end{bmatrix} \hspace{0.5cm} mod \hspace{0.5cm}  N.
\end{equation}

\noindent In equation~\ref{eq:arnoldcatmap}, $x, y \in \{0,1,...,N-1\}$, and $N$ is the size of host image. $a$  and $b$  are control parameters and increase the security in determining the mapping periodicity. $\begin{bmatrix} x \\y \end{bmatrix}$ is the main location of image pixels and $\begin{bmatrix} x^{'} \\ y^{'} \end{bmatrix}$ is the location of the mapped pixels. This mapping does not change the intensity of the image and only image data are disturbed. After several repetitions, the relationship between adjacent pixels is completely disturbed and the image looks distorted and meaningless~\cite{ref1}. Using this mapping as a pre-process in watermarking increases the security and reduces the possibility of targeted attacks. In the proposed method, the original image is Arnold mapped by Key1 (number of repetitions), 𝑎 and 𝑏 keys, and after disturbance of the pixels are subjected to the other algorithm phases. In addition to the security advantage that Arnold mapping provides, this mapping distributes the changes caused by embedding the watermark with a strong gain factor parameter in the entire image, which is not visually recognizable. However, using a strong power parameter without using Arnold mapping, the changes will be noticeable visually.
  
\subsection{Fast discrete curvelet transform (FDCuT)}\label{subsec.FDCuT}
\noindent One of the multi-scale transforms is the curvelet transform, which works better in distinguishing edges and curves compared to the other transform, and is more accurate to approximate and describe the dispersion and direction. Curvelet transform was first introduced on the basis of filtering the sub-bands and Ridgelet transform and is known as the first type curvelet. Due to defects in the first type curvelet, the second generation curvelet was presented based on the filtering the bypass in the Fourier domain. In the second generation curvelet, it initially takes a two-dimensional Fourier transform from the image, and then the image is fragmented into a series of discrete regions by a window in the frequency domain~\cite{ref2}. Then, the data are wrapped around the origin, finally the two-dimensional inverse Fourier transform is calculated on the wrapped data in order to calculate the curvelet coefficients.

Two types of fast discrete curvelet implementations are the Unequally Spaced Fast Fourier Transform USFFT-based curvelet and Wrapping-based curvelet. The first step in curvelet transform is decomposition of the signal into the sub-bands. These discrete transforms receive the Cartesian arrays (two-dimensional image) as $X[k1,k2], 0 \leq k1, k2 \leq k$ as the input and create the coefficients as the equation~\ref{eq:curvelet}.

\begin{equation}\label{eq:curvelet}
C^{D}(p,q,r)=\sum_{0\leq k1, k2\leq k}^{} X[k1,k2] \overline{\varphi^{D}_{p,q,r}(k1,k2)}.
\end{equation}

\noindent Where $\varphi^{D}_{p,q,r}$ is the wave-form digital curvelet, and $D$ represents the Digital word. The steps of discrete Wrapping curvelet transform briefly is as the following steps:

\begin{enumerate}[1),itemsep=0mm]
\item  Obtain Fourier coefficients $\hat{X}(k1, k2)$ by applying FFT.
\item Perform the following sub-steps for each $j$ scale and $i$ direction:
\begin{itemize}
\item Obtain the multiplication of $\hat{V}_{p,q}\hat{X}(k1, k2)$, where $\hat{V}_{p,q}$ is the parabolic window.
\item Wrap the multiplication around the origin and obtain $\tilde{X}{p,q}(k1,k2)=W(\hat{V}_{p,q}\hat{X})(k1,k2)$, where $0\leq k1 <2^{p}$ and $0\leq k2 <2^{\frac{p}{2}}$.
\item Calculate the discrete coefficients by applying the inverse FFT transform on the wrapped date, where, $C^{D}(p,q,r)$ is obtained.
\end{itemize}
\end{enumerate}

\begin{figure}
\center
\includegraphics[width=1\columnwidth]{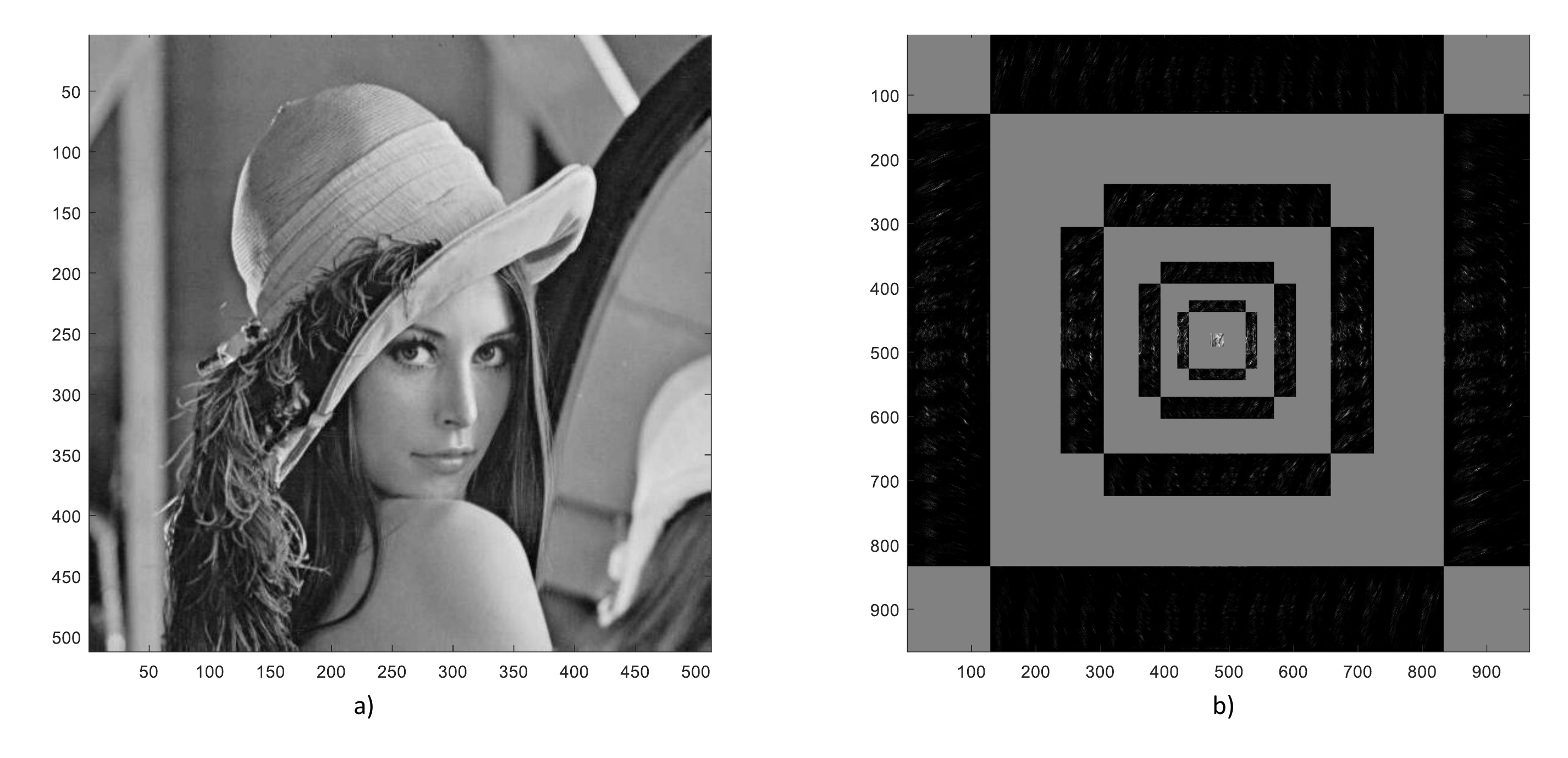}
\caption{a) The host image. b) decomposed image through FDCuT method.}
\label{fig:CurveletCoeff}
\end{figure}

\begin{figure}
\center
\includegraphics[width=1\columnwidth]{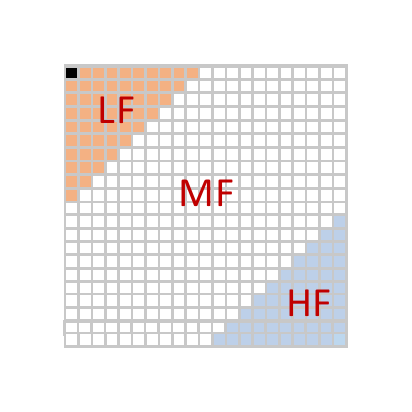}
\caption{LF, low frequency. MF, moderate frequency. HF, high frequency. Left and top corner, DC0 coefficient.}
\label{fig:DCT}
\end{figure}

\begin{algorithm}[!t]
 \KwData{$Seed$, $HF_{-}count$// Random generator seed and High Frequency coefficients count}
 \KwResult{ $SequenceOne$, $SequenceZero$ // Two Pseudo Random Noise corresponding to 0 and 1 bits of watermark using alphabet \{-1, 0, 1\}. }
 Set random generator seed to Seed \;
  $SequenceOne$ = $round(2\times(rand(1,HF_{-}count)-0.5))$\;
  $SequenceZero$ = $round(2\times(rand(1,HF_{-}count)-0.5))$\;

 \While{$((corr2(SequenceOne,SequenceZero) > 0.1)$ \&\& $ ((corr2(SequenceOne,SequenceZero)<-0.1)$}{
	  $SequenceOne$ = $round(2\times(rand(1,HF_{-}count)-0.5))$\;
	  $SequenceZero$ = $round(2\times(rand(1,HF_{-}count)-0.5))$\;
 }
 Return $SequenceOne$, $SequenceZero$ ;
 \caption{function GetPseudoNoisesFunc: Generate two pseudo random noise with low correlation}
      \label{algo.lowcorr}
\end{algorithm}

\begin{algorithm*}[t!]
\KwIn{$Watermark$ ($64$ bit), $Img$ (original image with size $512\times512$), $\textrm Key_1$, $\textrm Key_2$, $a$, $b$, $Gain$}
\KwOut{$W_{-}img$ (watermarked image)}
[$SequenceOne$, $SequenceZero$] = GetPseudoNoisesFunc($Key_1$);// Generate two pseudo random sequence with low correlation using Alg.~\ref{algo.lowcorr}.\\
$ImgMapped$ = ArnoldFunc($Img$, $Key_2$, $a$, $b$);//Apply Arnold Cat Map on the $Img$ with key $Key_2$,  $a$ and $b$.\\
$ImgDivided$ = DivideFunc($ImgMapped$);//Divide $ImgMapped$ into $64$ non-overlapping blocks of size $64\times64$ pixel. \\

$counter$=1;\\
 \While{$counter \leq 64$ }{
 
 $BlkApprox$ = FDCuTFunc($ImgDivided(counter)$);// Applying FDCuT transform on each block of $ImgDivided(counter)$ and get it's approximation sub-band ($21\times21$ pixel for each block).\\
$BlkDCT$ = DCT2Func($BlkApprox$);// Apply 2D-DCT transform on each $BlkApprox$.\\

  \If{$Watermark(counter) == 0$}{
  	$AugmenteNoiseZero$= $Gain * SequenceZero$;\\
	$BlkDCT$ = ReplaceNoiseFunc($AugmentedNoiseZero$);// Replace HF coefficients of $BlkDCT$ with $AugmentedNoiseZero$. \\		
   }
  \If{$Watermark(counter) == 1$}{
  	$AugmentedNoiseOne$ = $Gain * PequenceOne$;\\
	$BlkDCT$ = ReplaceNoiseFunc($AugmentedNoiseOne$);// Replace HF coefficients (Fig.~\ref{fig:DCT}) of $BlkDCT$ with $AugmentedNoiseOne$. \\		
   }
$BlkApprox$ = InverseDCT2Func($BlkDCT$);\\
$ImgDivided(counter)$ = InverseFDCuTFunc($BlkApprox$);\\    
$counter$ = $counter+1$; \\
 }
 $ImgMapped$ = AccumulateBlocksFunc($ImgDivided$);// Accumulation of embedded blocks to get a 512$\times$512 image.\\
 $W_{-}img$ = ArnoldFunc( $ImgMapped$, $period$-$Key_2$, $a$, $b$);// Apply Arnold Cat Map with key ($period$-$Key_2$).\\
 return $W_{-}img$;

\caption{Embedding algorithm }
\label{algo.embeding}
\end{algorithm*}

\begin{algorithm*}[t!]
\KwIn{$W_{-}Img$ (watermarked image ( Or attacked image) with size $512\times512$), $\textrm Key_1$, $\textrm Key_2$, $a$, $b$, $Gain$}
\KwOut{$Watermark$}
[$SequenceOne$, $SequenceZero$] = GetPseudoNoisesFunc($Key_1$);// Generate two pseudo random sequence with low correlation using Alg.~\ref{algo.lowcorr}.\\
$W_{-}ImgMapped$ = ArnoldFunc($W_{-}Img$, $Key_2$, $a$, $b$);//Apply Arnold Cat Map on the $W_{-}Img$ with key $Key_2$.\\
$W_{-}ImgDivided$ = DivideFunc($W_{-}ImgMapped$);//Divide $W_ImgMapped$ into $64$ non-overlapping blocks of size $64\times64$ pixel. \\

$counter$=1;\\
 \While{$counter\leq64$ }{
 
 $BlkApprox$ = FDCuTFunc($W_{-}ImgDivided(counter)$);// Applying FDCuT transform on each block of $W_{-}ImgDivided(counter)$ and get it's approximation sub-band ($21\times21$ pixel for each block).\\
$BlkDCT$ = DCT2Func($BlkApprox$);// Apply 2D-DCT transform on each $BlkApprox$.\\
$HFCoefficients$ = SelectHFFunc($BlkDCT$);// Select high frequency coefficients of $BlkDCT$.\\

$AugmentedNoiseZero$ = $Gain * SequenceZero$;\\
$AugmentedNoiseOne$ = $Gain * SequenceOne$;\\

Corr$_{-}$0 = Correlation($AugmentedNoiseZero$, $HFCoefficients$);\\
Corr$_{-}$1 = Correlation($AugmentedNoiseOne$, $HFCoefficients$);\\

  \If{Corr$_{-}0\leq Corr_{-}$1}{  	
		$Watermark(counter)=0$; \\
   }
 \If{Corr$_{-}1 < Corr_{-}$0}{  	
  		$Watermark(counter)=1$; \\
  }
  $counter$ = $counter+1$; \\
}
 return $Watermark$;

\caption{Extracting algorithm }
\label{algo.extracting}
\end{algorithm*}

\begin{figure*}[t]
  \center
  \includegraphics[width=1\textwidth]{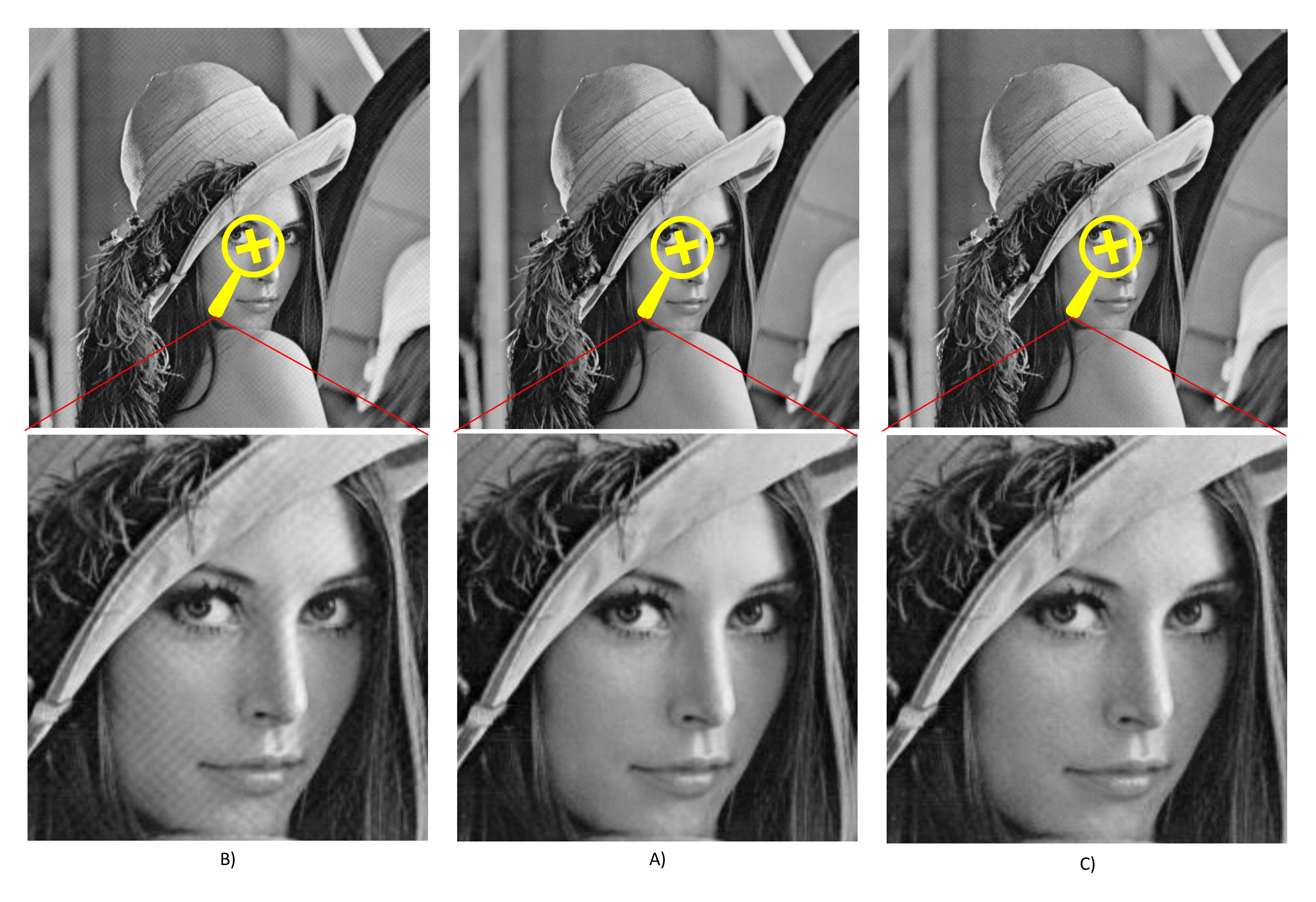}
  \caption{a) Original image. b) Watermarked image without Arnold Cat map (PSNR = 40.06 dB). c) Watermarked image using Arnold Cat map (PSNR = 40.13 dB).}\label{fig.ArnoldEffect}
\end{figure*}

\begin{figure*}[t]%
\centering
\begin{subfigure}{0.33\textwidth}
\includegraphics[width=\textwidth]{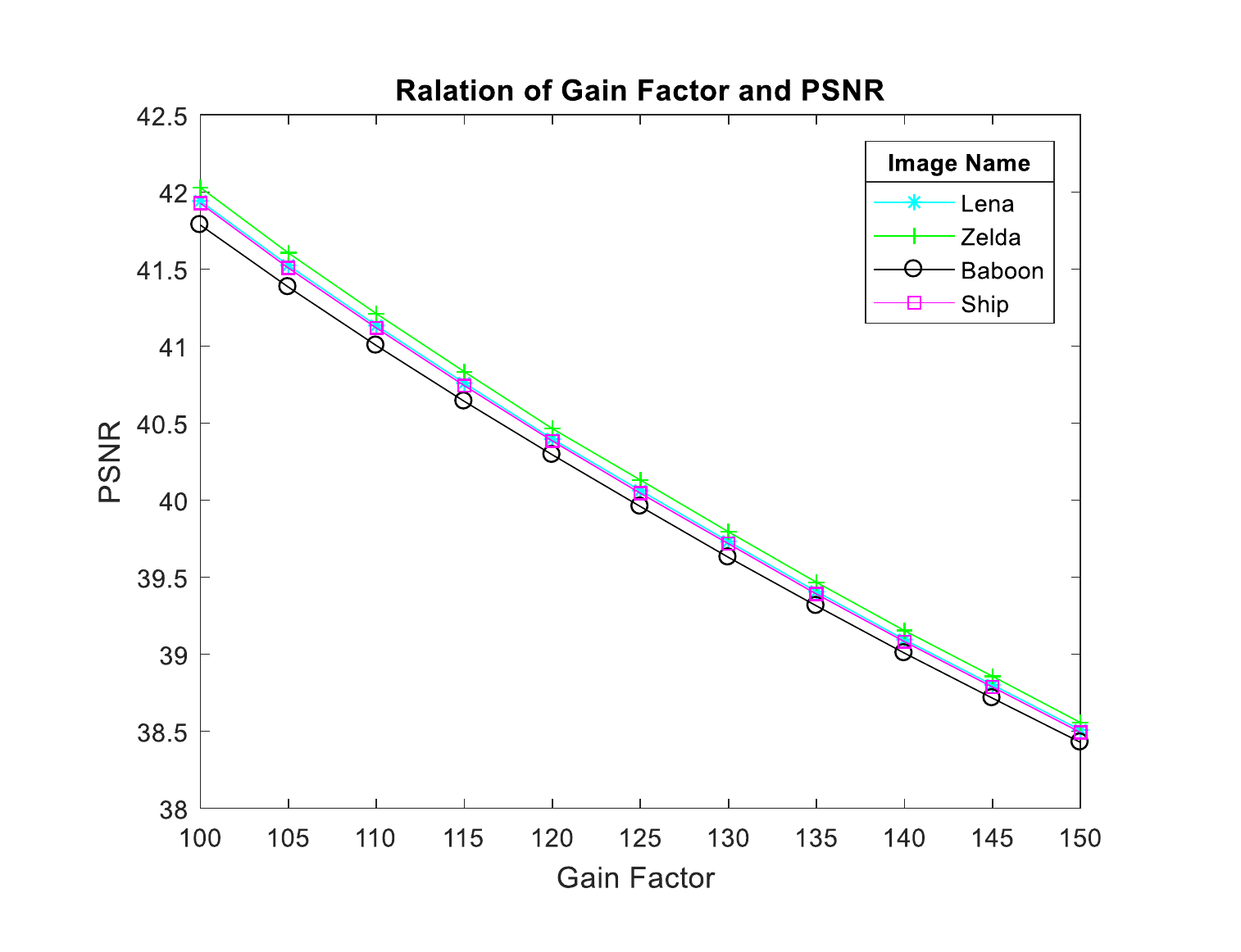}%
\caption{}%
\label{subfig.plots.a}%
\end{subfigure}\hfill%
\begin{subfigure}{.33\textwidth}
\includegraphics[width=\textwidth]{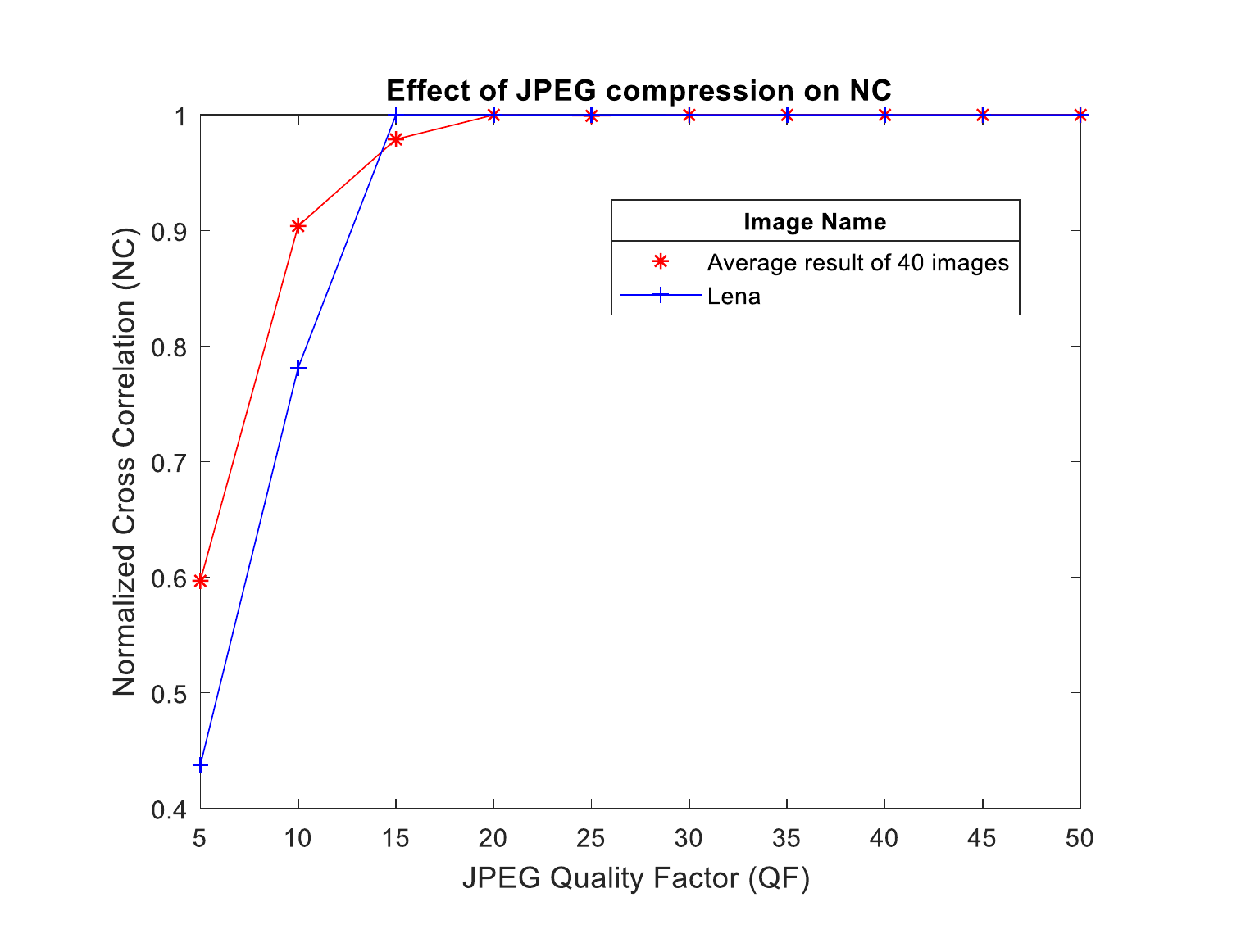}%
\caption{}%
\label{subfig.plots.b}%
\end{subfigure}\hfill%
\begin{subfigure}{.33\textwidth}
\includegraphics[width=\textwidth]{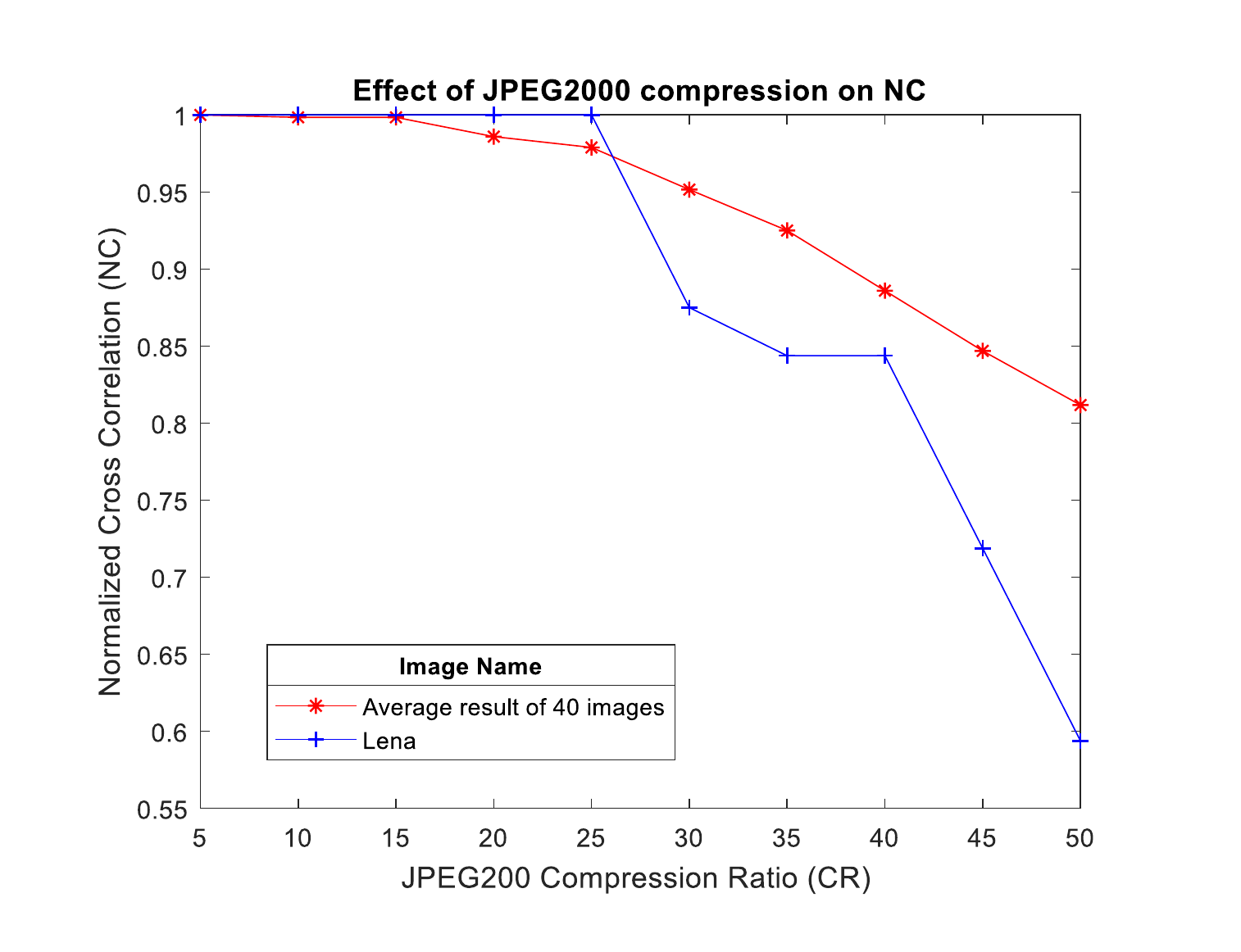}%
\caption{}%
\label{subfig.plots.c}%
\end{subfigure}\hfill%
\begin{subfigure}{.33\textwidth}
\includegraphics[width=\textwidth]{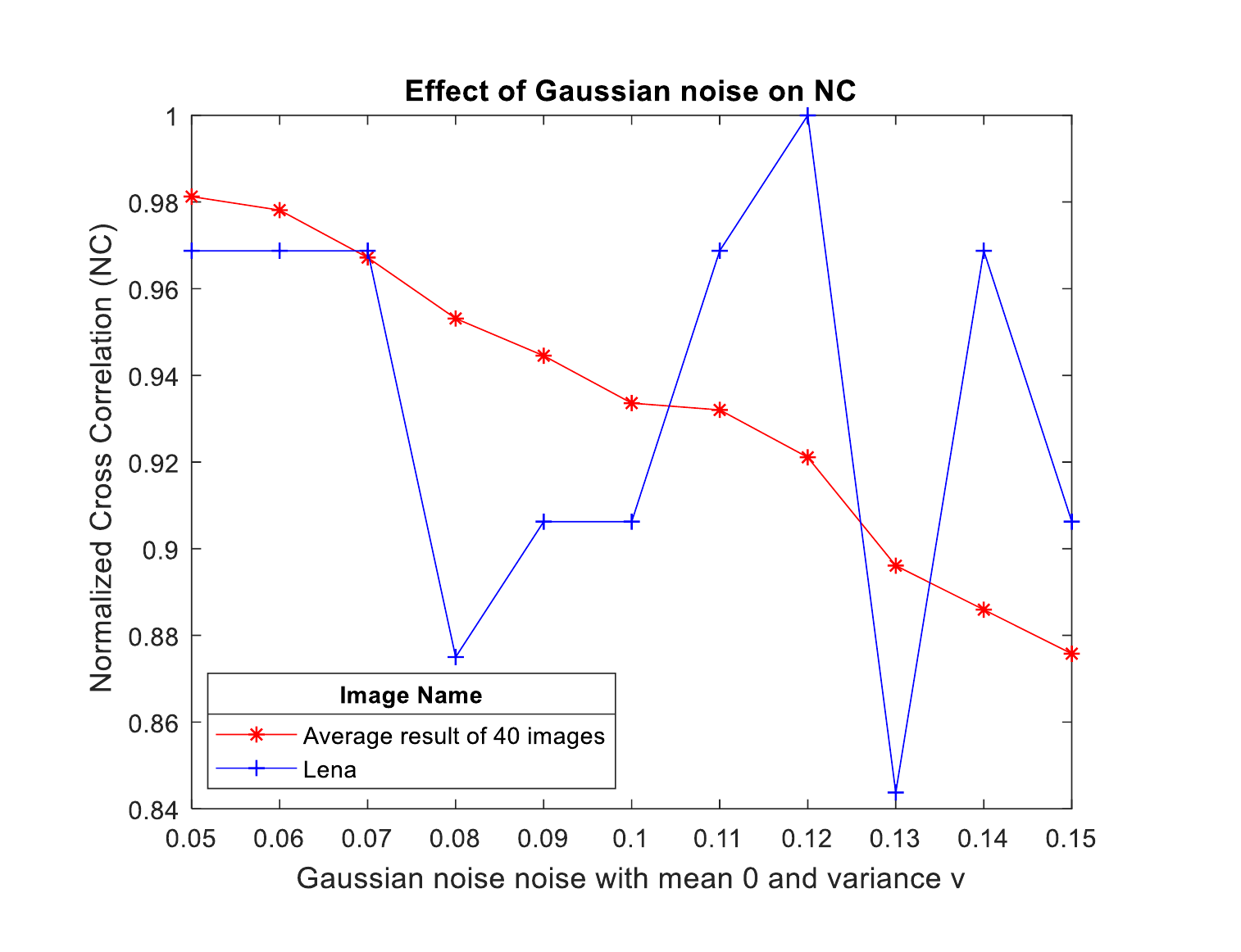}%
\caption{}%
\label{subfig.plots.d}%
\end{subfigure}\hfill%
\begin{subfigure}{.33\textwidth}
\includegraphics[width=\textwidth]{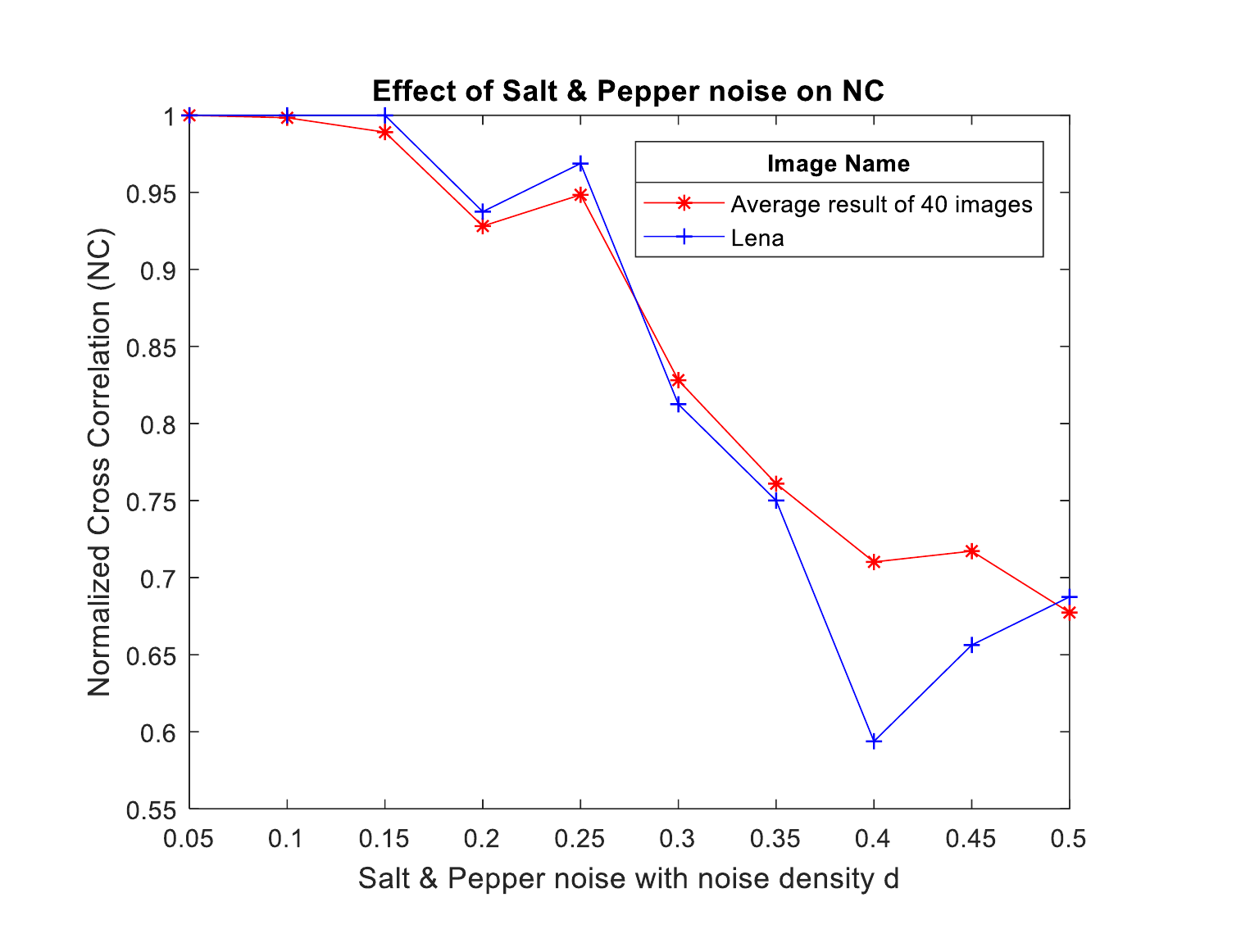}%
\caption{}%
\label{subfig.plots.e}%
\end{subfigure}\hfill%
\begin{subfigure}{.33\textwidth}
\includegraphics[width=\textwidth]{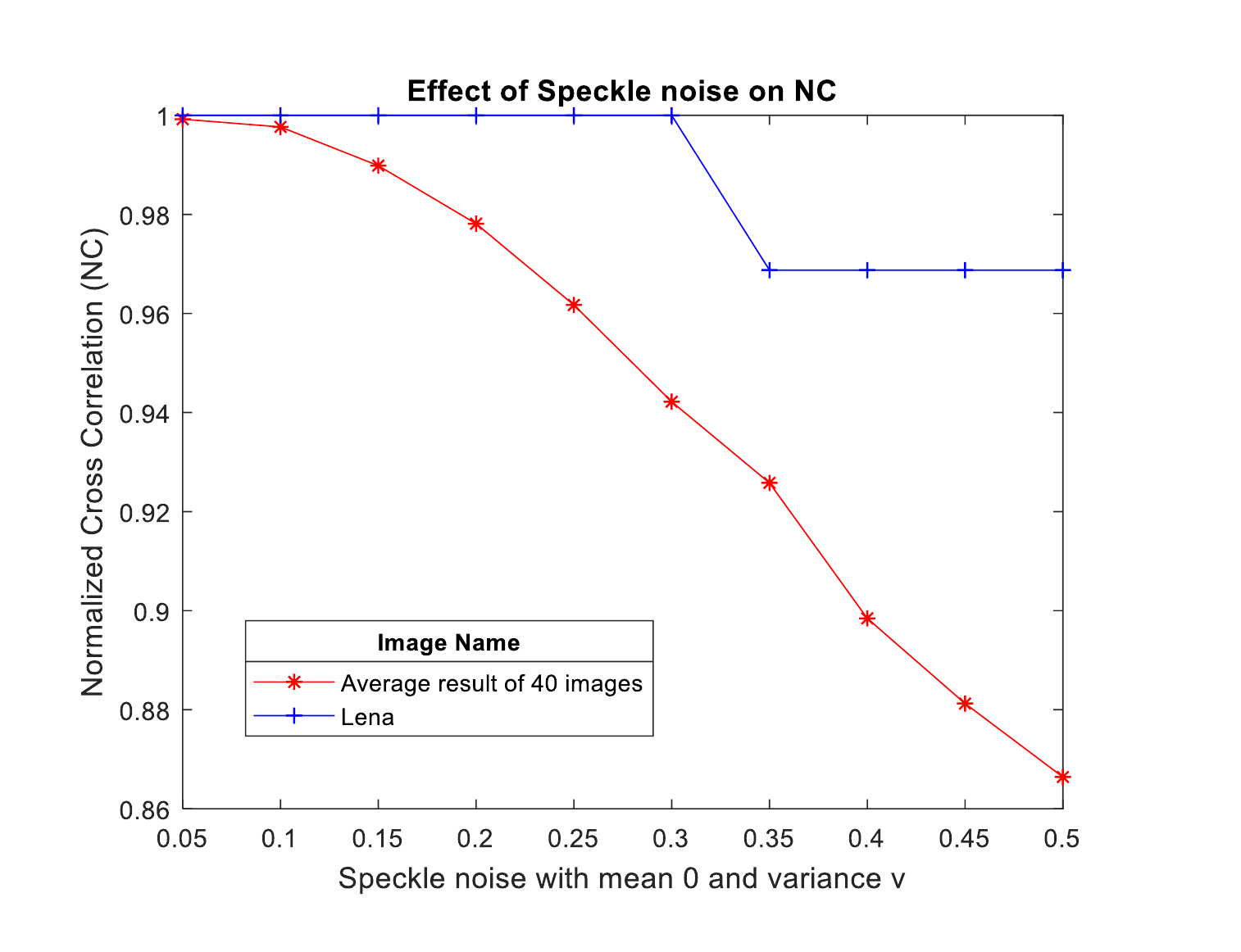}%
\caption{}%
\label{subfig.plots.f}%
\end{subfigure}\hfill%
\begin{subfigure}{.33\textwidth}
\includegraphics[width=\textwidth]{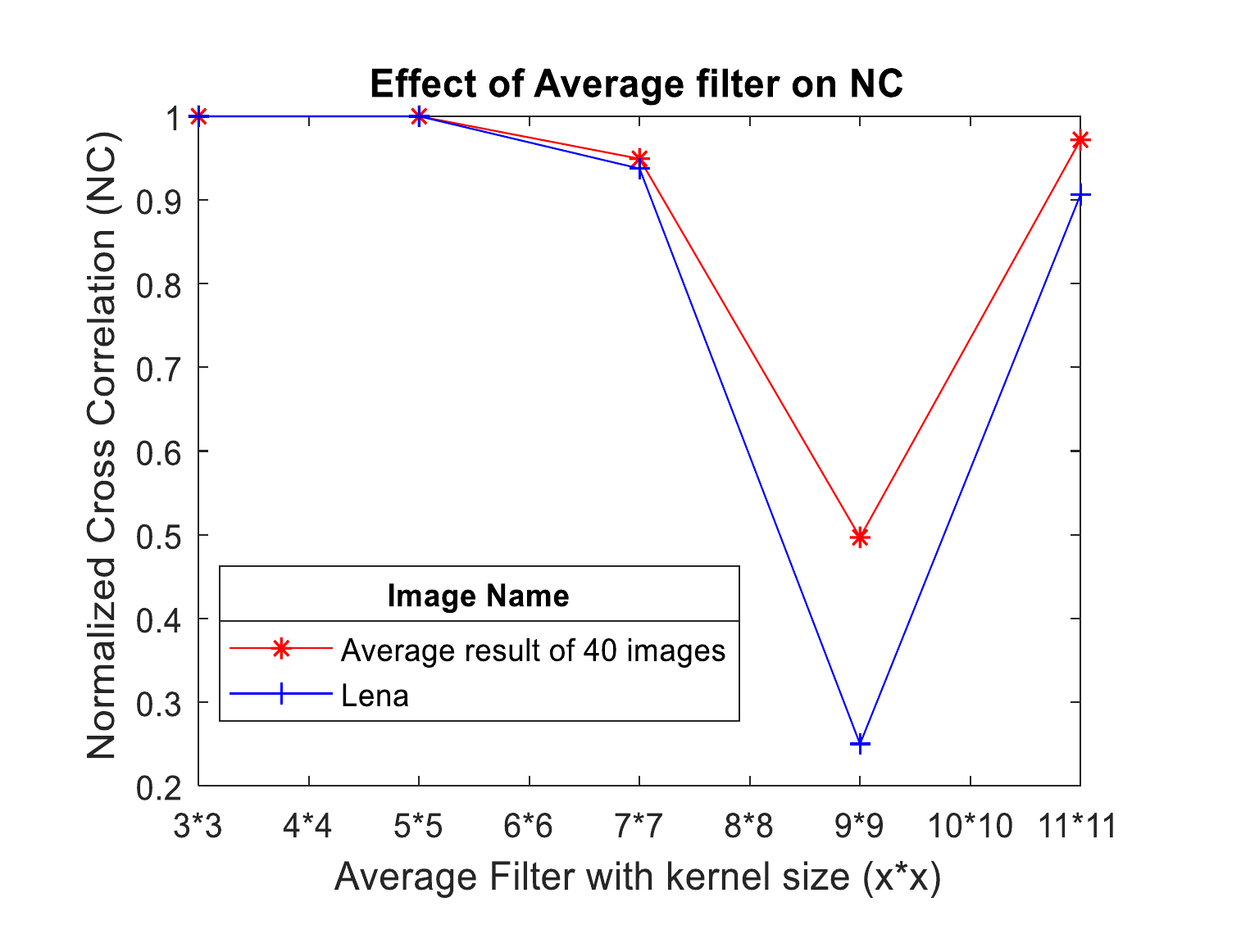}%
\caption{}%
\label{subfig.plots.g}%
\end{subfigure}\hfill%
\begin{subfigure}{.33\textwidth}
\includegraphics[width=\textwidth]{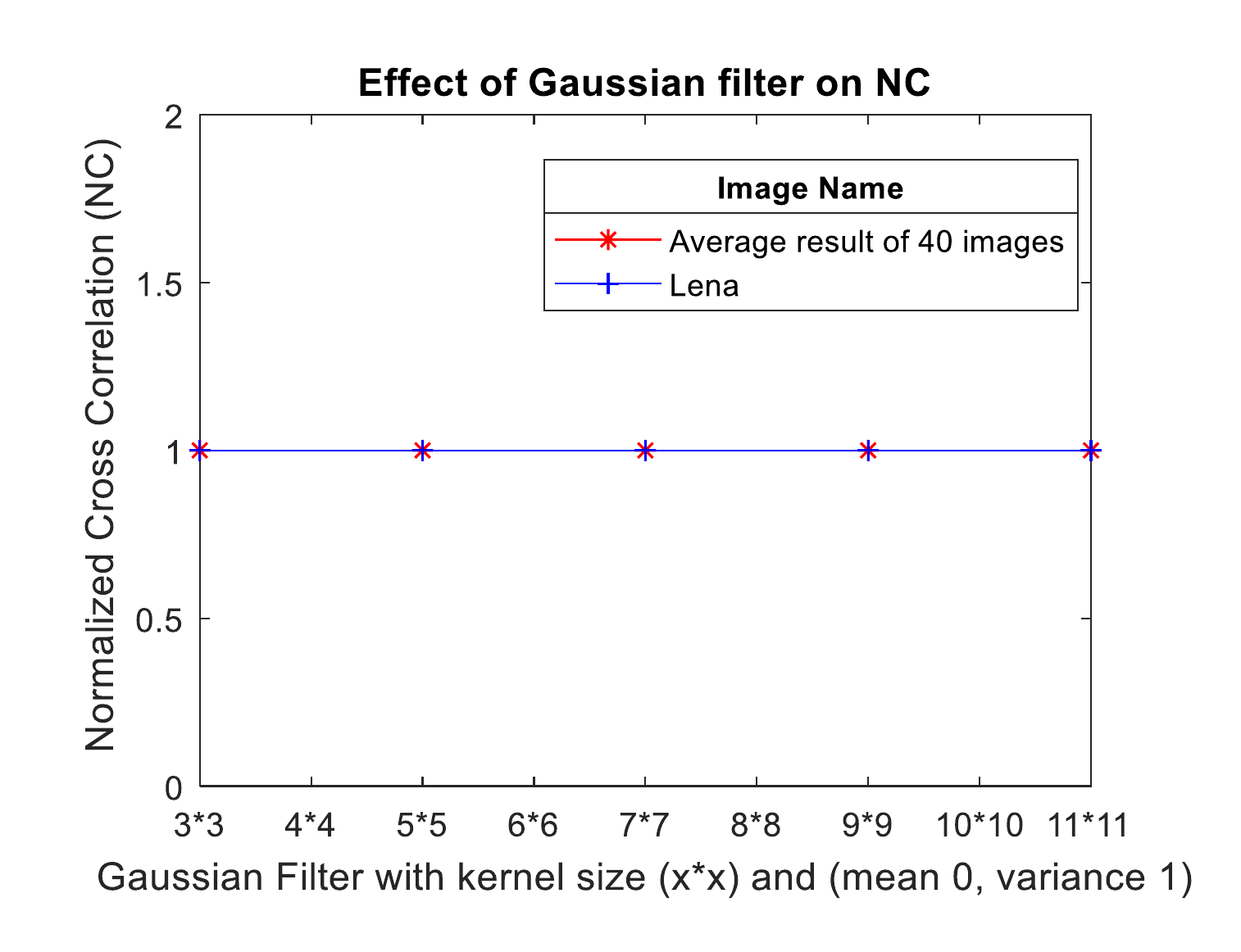}%
\caption{}%
\label{subfig.plots.h}%
\end{subfigure}\hfill%
\begin{subfigure}{.33\textwidth}
\includegraphics[width=\textwidth]{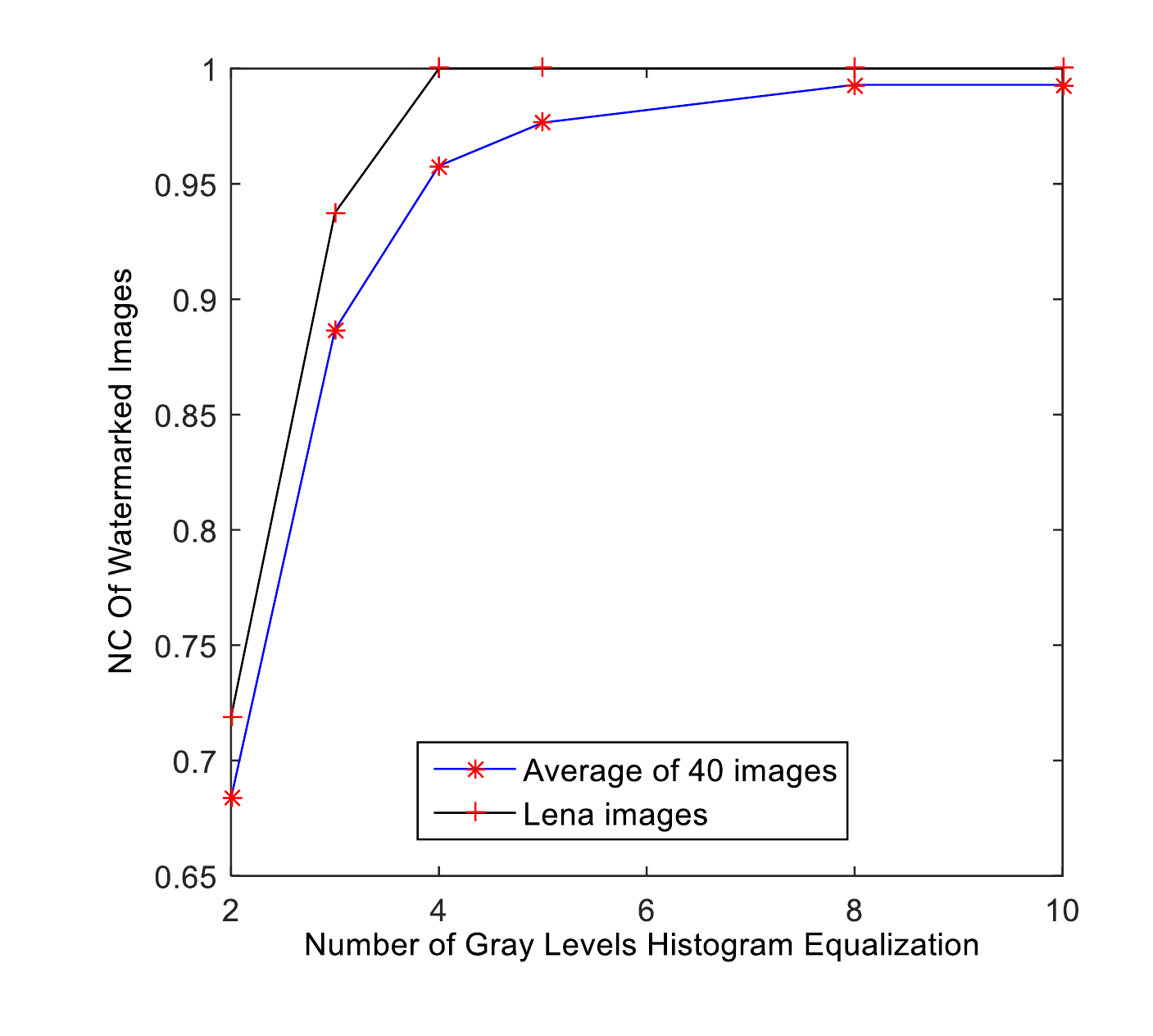}%
\caption{}%
\label{subfig.plots.i}%
\end{subfigure}\hfill%
\begin{subfigure}{.33\textwidth}
\includegraphics[width=\textwidth]{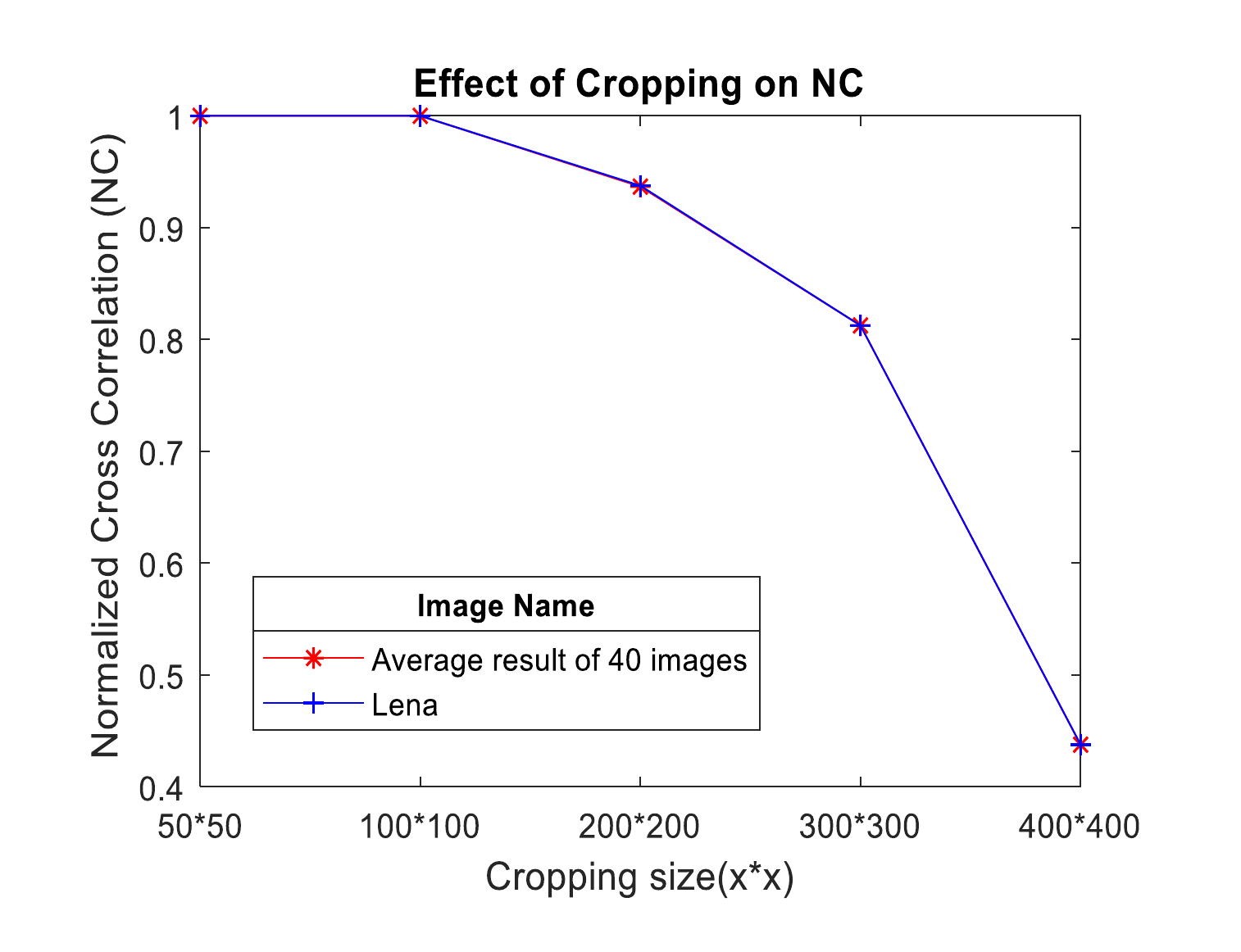}%
\caption{}%
\label{subfig.plots.j}%
\end{subfigure}\hfill%
\begin{subfigure}{.33\textwidth}
\includegraphics[width=\textwidth]{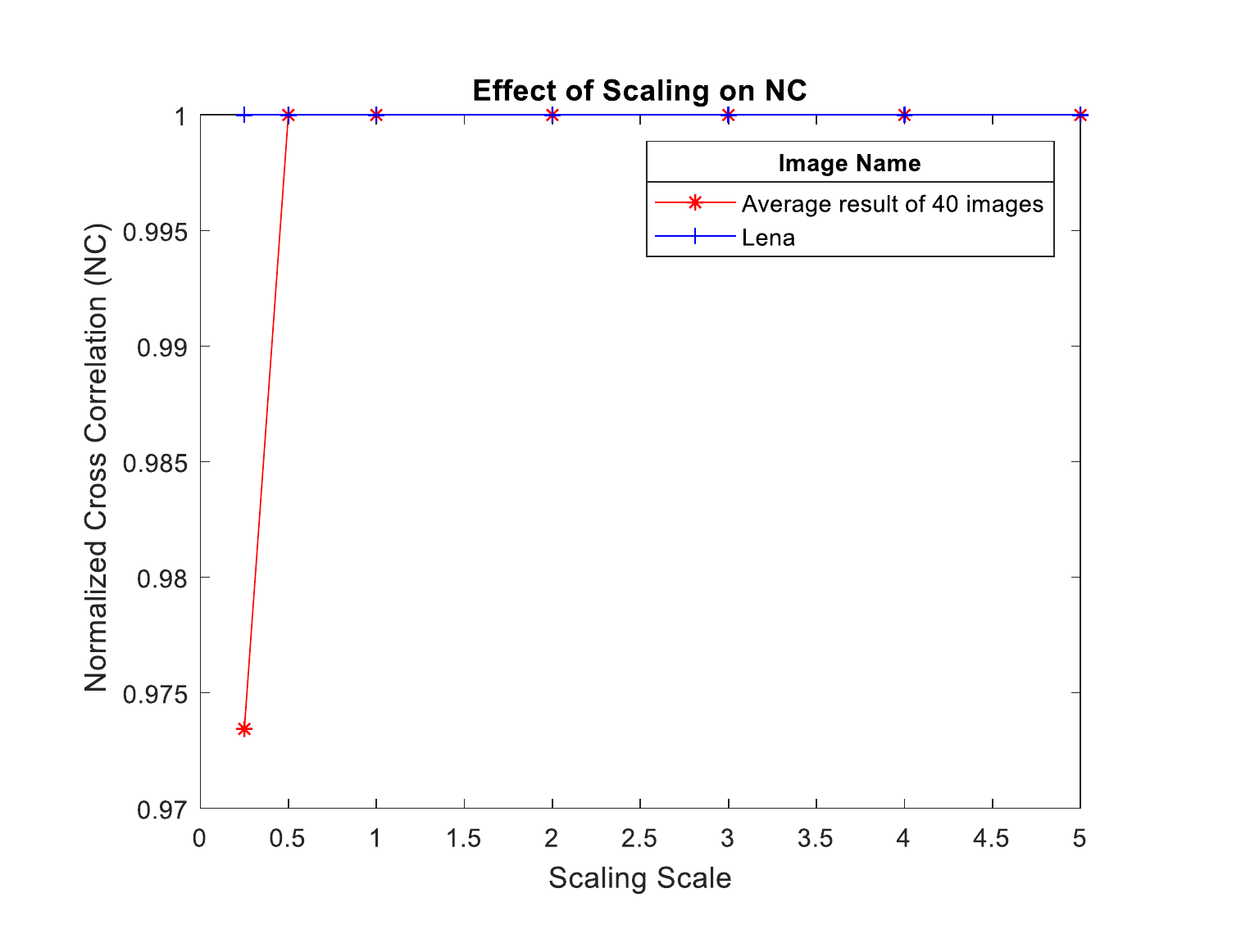}%
\caption{}%
\label{subfig.plots.k}%
\end{subfigure}\hfill%
\caption{Robustness of the proposed method against a number of common image processing attacks. Gain factor parameter is set 125 and the average quality of embedded images is more than 40 dB. Fig.~\ref{subfig.plots.a} shows comparison between gain factor and PSNR for 4 standard images. Fig.~\ref{subfig.plots.b} shows the NC in case of JPEG compression over Lena and 40 different watermarked images. Fig.~\ref{subfig.plots.c} shows the NC in case of JPEG2000 compression over Lena watermarked image. Fig.~\ref{subfig.plots.d} shows the NC in case of Gaussian noise over Lena and 40 different watermarked images. Fig.~\ref{subfig.plots.e} shows the NC in case of Salt \& Pepper noise over Lena and 40 different  watermarked images. Fig.~\ref{subfig.plots.f} shows the NC in case of Speckle noise over Lena and 40 different watermarked images.Fig.~\ref{subfig.plots.g} shows the NC in case of Average filter over Lena and 40 different watermarked images. Fig.~\ref{subfig.plots.h} shows the NC in case of Gaussian filter over Lena and 40 different watermarked images. Fig.~\ref{subfig.plots.i} shows the NC in case of Histogram Equalization over Lena and 40 different watermarked images. Fig.~\ref{subfig.plots.j} shows the NC in case of Cropping Lena and 40 different watermarked images. Fig.~\ref{subfig.plots.k} shows the NC in case of Scaling over Lena and 40 different watermarked images.}
\label{Fig.plots}
\end{figure*}

In Fig.~\ref{fig:CurveletCoeff}, the result of applying a wrap-based FDCuT is provided for a 512$\times$512 image. The FDCut transform divides the input image into a 21$\times$21 approximation sub-band, 4 moderate frequency sub-band with a 16-orientation parameter, and finally, a detail sub-band with the same size of the original image. The approximation sub-band ( 21$\times$21 pixels) of image is presented in the central part of the Fig.~\ref{fig:CurveletCoeff}-b. Medium frequency sub-bands and detail sub-band are presented respectively in black and gray. In the proposed method, the result of Arnold Cat mat is divided into 64$\times$64 non-overlapping blocks. Then, the FDCuT transform is applied on each block in order to get approximation sub-band. In the proposed method, approximation sub-band of each block is used for embedding just one bits of watermark. The reason for choosing the approximation sub-band for embedding the watermark is it's robustness against watermarking attacks and its appropriate distribution of information in the whole image.

\subsection{Discrete Cosine Transform (DCT)}\label{subsec:DCT}
\noindent Discrete cosine transform (DCT) is a method that transforms the spatial domain signal to frequency coefficients. DCT is one of the most prominent linear transforms, which is applicable in energy compression. The equation of discrete cosine transform is as the equation~\ref{eq:dct}:

\begin{equation}
\label{eq:dct}
\centering
\begin{multlined}
 C(u,v) =\alpha(u)\times\alpha(v)\times \\ \sum_{x=0}^{N-1}\sum_{y=0}^{N-1}
f(x,y)\cos\left ( \frac{\left ( 2x +1 \right )  u\pi}{2N} \right )\cos\left (\frac{\left ( 2y +1 \right )u\pi}{2N} \right );
\end{multlined}
\end{equation}
where
\begin{equation}
\alpha(u), \alpha(v)=\left\{\begin{matrix}
\sqrt{\frac{1}{N}}, &        u,v=0;
\\ \sqrt{\frac{2}{N}}, &  \hspace{1cm}u,v=1,2,\cdots,N-1.
\end{matrix}\right.\end{equation}

The low, moderate, and high frequencies are shown in Fig.~\ref{fig:DCT}, respectively, with LF, MF, and HF. Embedding information in LF coefficients provides the highest robustness against attacks such as JPEG compression. However, this will create the most destructive effect on the image. The embedding at the HF frequency has a relatively low robustness, but instead has a slight damage to the image. In this paper, robustness is provided because of the fact that the watermark information is embedded into the approximation sub-band calculated by FDCuT transform. However, in order to reduce the degradation effect of the embedding, the HF frequency coefficients of approximate sub-band is used to embed a bit of the watermark in each block.

\subsection{Weak correlation noises}\label{subsec:weakcorr}
\noindent In the proposed method, two random pseudo-noise strings using the alphabet {1, 0, -1} and to the number of coefficients in the HF frequency of each block (Fig.~\ref{fig:DCT}), are constructed as the symbol of the bit 0 and 1. Correlation of this two strings must be weak. The advantage of a weak correlation between the two randomly generated pseudo-noise strings shows itself at the extraction stage, since if the correlation not be weak, then due to the damage caused by the watermarking attacks on the image, the correlation of the two randomly embedded pseudo-noise strings may be close together and may disturb detection of 0 or 1 of the watermark. Therefore, as far as possible, correlation of two randomly generated pseudo-noise strings should be weak. The algorithm for producing two random pseudo-noise strings is as the Alg.~\ref{algo.lowcorr}.
In the Alg.~\ref{algo.lowcorr}, the Seed input is used to generate the same random values at the embedding and extraction stage. $HF_{-}count$ input represents the number of high-frequency coefficients of each block, to which two random pseudo-noise strings must be generated. $Corr2$ represents the function of calculating the correlation of the two randomly generated pseudo-noise strings, which its equation is as Eq.~\ref{eq.corr}. In Eq.~\ref{eq.corr}, $\bar{A}$ and $\bar{B}$ are the mean of the two $A$ and $B$ matrices.

\begin{equation}\label{eq.corr}
\begin{multlined}
corr(A, B)=\\ 
\left(\frac{\sum_{m}^{}\sum_{n}^{}(A_{mn}-\bar{A})(B_{mn}-\bar{B})}{\sqrt{\left(\sum_{m}^{}\sum_{n}^{}(A_{mn}-\bar{A})^2\right) \left(\sum_{m}^{}\sum_{n}^{}(B_{mn}-\bar{B})^2\right)}}\right).
\end{multlined}
\end{equation}

\subsection{Watermark Embedding}\label{subsec:embedding}
\noindent As stated in the previous sub-sections, the final location of the embedding watermark is the high-frequency coefficients generated from each 21$\times$21 block, as shown in Fig.~\ref{fig:DCT}. The value of these coefficients is close to zero, and changing them does not cause much side effects in the original image. In the proposed method, in each 21$\times$21 block, only one bit of watermark is embedded. Therefore, based on the 0 or 1 bit of the watermark, one of the two random pseudo-noise strings must be embedded. The embedding method used in this paper is replacement method. In other words, the result of the multiplication of random pseudo-noise strings in a relatively large gain factor, the random amplified pseudo-noise strings are obtained and then one of the pseudo-noise strings is replaced in the high-frequency coefficients of each block proportional to 0 or 1 bit of the watermark. The gain factor used in this embedding is much larger than the initial coefficient value and causes a change in the quality of the embedding image. However, as a result of using Arnold Cat map to create a disturbance in the pixel location, the changes from the random amplified pseudo-noise string will not be visually recognizable. The embedding algorithm of the 64-bit watermark in the host image is shown in Alg.~\ref{algo.embeding}.

\subsection{Watermark Extraction}\label{subsec.extraction}
\noindent After embedding the watermark in the host image, the embedded image may be intentionally or unintentionally attacked, and the extracted watermark is not exactly the same as the embedded watermark. The main parts in the extraction phase are similar to those in the watermark embedding steps. This means that at first the Arnold Cat map must be applied on the image by the keys similar to the embedding stage, and then divided into non-overlapping 64$\times$64 blocks. Then, FDCuT and DCT transforms should be applied to each block to calculate the HF coefficients of each block. An important stage to be followed is to decide on the bit embedded in the desired block. To do this, the correlation between the calculated HF coefficients with the two randomly amplified pseudo-noise strings is calculated for both of the 0 and 1 bit of the watermark. If the correlation of the HF coefficients with the amplified random pseudo-noise string related to the bit 0 is greater than the amplified random pseudo-noise string of bit 1, it indicates that the bit 0 was embedded, otherwise it indicates that the bit 1 was embedded. The watermark extraction algorithm is shown in algorithm 3.

\section{Experimental result}
\noindent In these experiments, 40 standard images of the USC-SIPI image database were used~\cite{ref21}. Many of these images such as Lena, Zelda, Baboon, Camera Man, Pepper, etc. have been used repeatedly in watermarking and steganography as a test image. To create different attacks, the Matlab software was used to evaluate the proposed method and attempts were made to provide the more detailed explanations of the attack on the plots. The criterion referred in this paper and most articles is the peak signal-to-noise ratio (PSNR). This criterion indicates the amount of noise added to the image by the watermark insertion in it. It is also used in image recovery techniques to evaluate the quality of the extracted image. Although this parameter does not exactly imply the invisibility of the visual watermark in the image, it provides a proper algebraic relationship for the optimal amount of changes in the image. The definition of this evaluating criterion is presented as the equation~\ref{eq.psnr}.

\begin{equation}\label{eq.psnr}
  \textrm{PSNR}(f,f_w)=
  10\log_{10}\left [{\frac{\max_{\forall(m,n)}f^2(m,n)}{\frac{1}{N_f}\sum_{\forall(m,n)}{\left( f_w(m,n)-f(m,n) \right)}^2}}\right].
\end{equation}

Equation~\ref{eq.psnr}, represents the value of the PSNR in decibel unit. In this equation, $f$ is the original image, $f_{w}$ the watermarked image, $(m, n)$ the index of the images pixels, and $N_f$ is the number of pixels in each  $f$  and $f_{w}$ images. The larger values of this criterion shows the imperceptibility of watermarking method. Regularly, the values of about 40 dB are acceptable values for this criterion in image watermarking~\ref{eq.psnr}.

One of the most commonly used criteria for evaluating the extracted watermark is normalized cross-correlation (NC). The definition of this parameter is as the equation~\ref{eq.nc}.

\begin{equation}\label{eq.nc}
  NC = \left (
  \frac{{\sum_{j=1}^{N_w}{W(i,j)\times \acute{W}(i,j)}}}{\sum_{i=1}^{N_w}{{W(i,j)}}}\right).
\end{equation}

In equation~\ref{eq.nc}, $W$ and $\acute{W}$ are respectively the embedded watermark and extracted watermark, and $N_w$ is the size of the watermark bits. The value of the watermark is assumed within $\{-1, 1\}$, and the NC value will be 1 if there is no error in the extracted watermark. In fact, the closer this criterion to 1, the watermarking method is more robust~\cite{ref5}.

The Gain factor parameter in this paper is 125 since, with this Gain factor, the average visual quality of the tested images is above 40 dB. As it is observed in Fig.~\ref{fig.ArnoldEffect}-b, with such a Gain factor, there are noticeable changes in the watermarked image. However, in Fig.~\ref{fig.ArnoldEffect}-c, where Arnold Cat map was used, the changes made in the watermarked image are not visibly observable and tangible. In fact, Arnold Cat map distributes the changes in the entire of image.

In Fig.~\ref{Fig.plots}, the output of the proposed method is shown on the 40 selected images in the presence of common image processing attacks. Furthermore, considering the fact that the standard Lena image is important from the standpoint of watermarking articles, the proposed method robustness to this image is shown separately into the plots. As it is observed, the proposed method has a high resistance to all kinds of noise, resizing and JPEG2000 compression. It should be noted that the compression rate used in the JPEG2000 attack indicates that how much the compressed image size is smaller than the original image. For example, the compression rate of 10 means that the compressed image will be a tenth of the original image size. Based on the results of Figure 5, it is concluded that the proposed method is robust against JPEG compression with a quality factor of 10, however, for JPEG compression with a quality factor below 10, the watermark is severely degraded. Additionally, the proposed method is capable of handling the resizing attack to the image to a quarter of the original size easily, however, if the image size is less than a quarter of the original size, the watermark will be degraded and cannot be extracted.

The comparison of the proposed method results with four robust methods presented in recent years is shown in Tables~\ref{tbl.comp1}~and~\ref{tbl.comp2}. For this comparison, the same images are used. Moreover, the Gain factor parameter is set to match the quality of the embedded image obtained by the proposed method and the compared methods. In this comparison, it was attempted to evaluate the same attacks on the watermarked images. As it can be seen, the proposed method in most cases has more robustness to attacks than the compared methods, however, the robustness of the proposed method against JPEG compression attack is at the same level of the compared methods or less.

\begin{table*}[b]
,\caption{Comparison of proposed method with \cite{ref4,ref6}}\label{tbl.comp1}
  \centering
  \begin{tabular}{l c c | l c c }
  \hline
  \multicolumn{3}{c|}{Lena image} &  \multicolumn{3}{c}{Lena image}  \\
  \hline
  \rotatebox[origin=c]{90}{Attack} & \rotatebox[origin=c]{90}{Proposed method    (PSNR=40.10)} & \rotatebox[origin=c]{90}{\cite{ref4}(PSNR=40.40)}&  \rotatebox[origin=c]{90}{Attack} & \rotatebox[origin=c]{90}{Proposed method     (PSNR=46.36)} & \rotatebox[origin=c]{90}{\cite{ref6}(PSNR=46.18)}\\
  \hline

  JPEG QF=10 & 0.71 & \textbf{0.93}
   & Median[3$\times$3] &  \textbf{1.0}  & 0.95
  \\
  JPEG QF=20 & \textbf{1.0} & 1.0  & Average[3$\times$3] &  \textbf{1.0}  & 0.75 \\
  JPEG QF=30 &   \textbf{1.0} & \textbf{1.0} & JPEG QF=10 &  0.53  & \textbf{0.92} \\
  Scaling 1/4 & \textbf{1.0} & \textbf{1.0}  & Cropping 10\% &  \textbf{0.96}  & 0.86  \\
  Cropping 1/4 & 0.81  &  \textbf{0.90} & Gaussian Noise 0.01 &  \textbf{0.96}  & 0.85  \\
  Average[3$\times$3] &  \textbf{1.0}  & \textbf{1.0} & Salt \& Pepper noise 0.1 &  \textbf{0.87}  & 0.85  \\
  Average[5$\times$5] &  \textbf{1.0}  & 0.81 \\
  Average[7$\times$7] & \textbf{93.0}  &  0.46   \\
 JPEG2000 QF=10 &  \textbf{1.0}  & \textbf{1.0} \\
 JPEG2000 QF=20 &  \textbf{1.0}  & 0.81  \\
 JPEG2000 QF=30 &  \textbf{0.84}  & 0.59  \\
  Median[3$\times$3] &  \textbf{1.0}  & \textbf{1.0}  \\
  Median[5$\times$5] &  \textbf{1.0}  & \textbf{1.0}  \\
  Median[7$\times$7] &  0.62 & \textbf{0.81}  \\
  Gaussian noise 0.10 &  \textbf{0.93}  & 0.81  \\
  Gaussian noise 0.15&  \textbf{0.88}  & 0.31  \\
  Salt \& Pepper noise 0.30 &  \textbf{0.81}  & 0.65  \\
  Salt \& Pepper noise 0.35 &  \textbf{0.75}  & 0.62  \\
  Speckle noise 0.30 &  \textbf{1.0}  & 0.62  \\
  Speckle noise 0.35 &  \textbf{0.96}  & 0.53  \\
  \hline
\end{tabular}
\end{table*}

\begin{table*}
,\caption{Comparison of proposed method with \cite{ref7,ref8}}\label{tbl.comp2}
  \centering
  \begin{tabular}{l c c | l c c }
  \hline
  \multicolumn{3}{c|}{Lena image} &  \multicolumn{3}{c}{Lena image}  \\
  \hline
  \rotatebox[origin=c]{90}{Attack} & \rotatebox[origin=c]{90}{Proposed method    (PSNR=41.1)} & \rotatebox[origin=c]{90}{\cite{ref7}(PSNR=41.94)}&  \rotatebox[origin=c]{90}{Attack} & \rotatebox[origin=c]{90}{Proposed method     (PSNR=45.46)} & \rotatebox[origin=c]{90}{\cite{ref8}(PSNR=45.39)}\\
  \hline

  JPEG QF=10 & 0.71 & \textbf{0.87} & JPEG QF=10 & 0.62 & \textbf{0.63}
  \\
  JPEG QF=20 & \textbf{1.0}  & 0.94 & JPEG QF=20 &  \textbf{0.87}  & 0.85\\
  JPEG QF=30 & \textbf{1.0}  & 0.99 & JPEG QF=30 &  \textbf{0.96}  & 0.92 \\
  JPEG QF=40 & \textbf{1.0}  &  \textbf{1.0} & JPEG QF=40 &  0.96  & \textbf{0.97}  \\
  JPEG QF=50 &  \textbf{1.0}  &  \textbf{1.0} & JPEG QF=50 &  \textbf{1.0}  & 0.99 \\
  JPEG QF=60 &  \textbf{1.0}  &  \textbf{1.0} & JPEG QF=60 &  \textbf{1.0}  &  \textbf{1.0}  \\
  Scaling 1/4 & \textbf{1.0}  & 0.98 & Median[3$\times$3] &  \textbf{0.96}  & 0.93  \\
  Cropping 1/4 & 0.81  &  \textbf{0.93} & Median[5$\times$5] &  0.81  & \textbf{0.86}  \\
  Average[3$\times$3] &  \textbf{1.0}  & 0.89 & Scaling 1/2 &  \textbf{1.0}  & 0.96  \\
  Average[5$\times$5] &  \textbf{1.0}  & 0.73 & Scaling 1/4 &  \textbf{1.0}  & 0.97  \\
  Average[6$\times$6] & \textbf{1.0}  &  0.64 & Crop 25\% &  0.81  &  \textbf{0.88}  \\
  Histogram Eq &  \textbf{1.0}  & 0.92  & Gaussian Filter[3$\times$3] &  \textbf{1.0}  &  0.99 \\
  Median[3$\times$3] &  \textbf{1.0}  & 0.95  \\
  Median[5$\times$5] &  \textbf{1.0}  & 0.81  \\
  Median[6$\times$6] &  \textbf{0.75} & 0.62  \\
  Gaussian noise 0.02 &  \textbf{0.93}  & 0.85  \\
  Salt \& Pepper noise 0.01 &  \textbf{1.0}  & 0.75  \\
  Speckle noise 0.01 &  \textbf{1.0}  & 0.74  \\
  \hline
\end{tabular}
\end{table*}

\section{Conclusion and future work}
\noindent In this paper, a robust and blind image watermarking technique was proposed that has the ability to extract embedded watermark after strong attacks such as intense noise, image compression, and image quality enhancements processing. In this method, two pseudo-noise strings with weak correlation are constructed as the symbol of each bit 0 and 1 of the watermark. The embedding location in this method is the high frequency coefficients of the approximation sub-band of FDCuT transform. Arnold Cat map is used to enhance the security and imperceptibility of embedding. Using Arnold Cat map as a preprocessing in this method provides the possibility of amplifying two pseudo-noise strings as possible, thereby enhancing the robustness of the proposed method. In order to more accurately evaluate and compare the proposed method with recent robust methods, it was tried to perform the comparison in the same test conditions, such as the same image and the same embedded image quality. Comparison of the proposed method with recent robust methods indicates the high robustness of the proposed method.

As the future work, the proposed method can be evaluated on colorful images, and different sub-bands and features of the colorful images can be used to increase the robustness of the proposed method. Some methods such as SIFT and SURF can also be used to find the key points and embed the watermark in those locations to make the proposed method robust against rotation attacks.

\section*{References}
\bibliographystyle{elsarticle-num}

\end{document}